\documentclass[prl,english,twocolumn,amsfonts,amssymb,floats,aps]{revtex4}

\usepackage[final,dvips]{epsfig}
\usepackage{amsmath,amsfonts}
\usepackage{bbm}
\usepackage{float}
\usepackage[caption = false]{subfig}
\usepackage{graphicx}
\usepackage{color}
\usepackage{hyperref}
\usepackage[normalem]{ulem}

\newcommand{\jp}{J} 
\newcommand{\jn}{K} 
\newcommand{\jnx}{K^x} 
\newcommand{\jny}{K^y} 
\newcommand{\jnz}{K^z} 
\newcommand{\ani}{\alpha} 
\newcommand{\dm}{D} 
\newcommand{\gk}{\gamma'_{\vec{k}}} 

\newcommand{\kk}{\vec{k}} 
\newcommand{\ii}{\iota} 
\newcommand{\zt}{\mathbbm{Z}_{2}} 
\newcommand{\po}{\hat{P}} 
\newcommand{\ch}{\hat{\mathcal{C}}} 

\begin{document}

 \title[]
{
$\mathbbm{Z}_{2}$ topological quantum paramagnet on a honeycomb bilayer
}
\author{Darshan G. Joshi}
\email[]{d.joshi@fkf.mpg.de}
\affiliation{Max-Planck-Institute for Solid State Research, D-70569 Stuttgart, Germany}

\author{Andreas P. Schnyder}
\email[]{a.schnyder@fkf.mpg.de}
\affiliation{Max-Planck-Institute for Solid State Research, D-70569 Stuttgart, Germany}

\date{\today}

\begin{abstract}

Topological quantum paramagnets are exotic states of matter, whose magnetic excitations have a topological band structure, while
the ground state is topologically trivial. 
Here we show that a simple model of quantum spins on a 
honeycomb bilayer hosts a time-reversal-symmetry protected 
$\mathbbm{Z}_2$ topological quantum paramagnet ({\em topological triplon insulator}) in the presence of spin-orbit coupling. 
The excitation spectrum of this quantum paramagnet consists of three triplon bands, two of which carry a nontrivial $\mathbbm{Z}_2$ index.
As a consequence, there appear two counterpropagating triplon excitation modes at the edge of the system.
We compute the triplon edge state spectrum and the $\mathbbm{Z}_2$ index for various parameter choices.
We further show that upon making one of the Heisenberg couplings stronger, the system undergoes a topological quantum phase transition, 
where the $\mathbbm{Z}_2$ index vanishes, to a different topological quantum paramagnet. In this case the counterpopagating triplon edge modes are disconnected from the bulk excitations and
are protected by a  chiral and a unitary symmetry.
We  discuss possible realizations of our model in real materials, in particular d$^{4}$ Mott insulators, and their potential applications.
\end{abstract}


\maketitle


\emph{Introduction.--} 
The topology of quasiparticle band structures is of great interest for  fundamental science and possible technological applications~\cite{hasan_kane,classif2,bansil_RMP_16,armitage_mele_RMP_18}.
Not only fermionic but also bosonic quasiparticles can exhibit topological band structures. 
This has been demonstrated in a number of  artificial systems, such as for electromagnetic waves in dielectric superlattices~\cite{wang_joannopoulos_photonics,Lu622_science_Weyl_photonic_crystal}
or for polaritons in microcavities~\cite{jean_microcavity_polariton_Nat_photon_17}.
Bosonic quasiparticles with topological properties can also arise intrinsically in a variety of materials, e.g., as topological phonons in systems with isostatic lattices~\cite{kane_lubensky_NatPhys_13}
 as topological spin excitations in quantum magnets~\cite{owerre_sci_rep_17,joshi_PRB_17,sk_kim,owerre_PRB_16,owerre1, owerre2,baolong_shindou_PRB_16,kh_joshi,kh_penc,balents_gang_weyl_magnon,dirac_magnons_PRL_17,top_magnons_AF_arXiv_17},
or as topological triplon bands in dimerized magnets~\cite{ss_hall}, which have been observed experimentally~\cite{ss_hall_exp}.

The study of topological spin excitations is enjoying growing activity, both due to its fundamental importance and its potential relevance for magnonic devices~\cite{magnonics}. 
For example, topological magnon~\cite{sk_kim,owerre_PRB_16,owerre1,owerre2,baolong_shindou_PRB_16,kh_joshi,kh_penc} and triplon insulators~\cite{ss_hall,ss_hall_exp}, as well as Dirac~\cite{dirac_magnons_PRL_17,top_magnons_AF_arXiv_17} and Weyl magnon semimetals~\cite{balents_gang_weyl_magnon} have been investigated. The magnon and triplon bands in these quantum magnets carry a nonzero Chern number, which by the bulk-boundary correspondence,
gives rise to chiral magnon and triplon modes at the surface. Since these chiral surface modes carry spin with low dissipation and are protected against disorder, they could be
utilized as efficient channels for spin transport~\cite{ruckriegel_PRB_18}. 
However, in contrast to electronic topological insulators, the chiral surface magnons and triplons are excited states with an
energy considerably higher than the bulk Goldstone modes of the ordered magnet. Hence, due to coupling to the low-energy bulk modes,
 these topological surface magnons and triplons are strongly damped~\cite{topo_decay}, which suppresses the surface spin transport. 

Recently, it was shown that topological spin excitations can also exist in the quantum-disordered paramagnetic phase of a spin ladder~\cite{joshi_PRB_17}.
This one-dimensional topological quantum paramagnet exhibits protected triplon end states.
In contrast to the magnon~\cite{sk_kim,owerre_PRB_16,owerre1,owerre2,baolong_shindou_PRB_16,kh_joshi,kh_penc,dirac_magnons_PRL_17,top_magnons_AF_arXiv_17,balents_gang_weyl_magnon} and triplon~\cite{ss_hall,ss_hall_exp} surface states of the aforementioned ordered magnets, the triplon end states of the quantum-disordered paramagnet~\cite{joshi_PRB_17} are only weakly damped due to energy-momentum constraints and the absence of Goldstone modes.
For applications it would be advantageous to have a \emph{two-dimensional} version of this quantum paramagnet, with protected triplon edge states forming a robust channel for spin transport. 

In this paper, we provide an example of such a two-dimensional topological quantum paramagnet.
 We consider a spin-1/2 system of two coupled honeycomb layers, with  strong antiferromagnetic exchange interactions between the layers
 and weaker intralayer Heisenberg and Dzyaloshinskii-Moriya (DM)  interactions.
 The dominant interlayer antiferromagnetic exchange leads to a coupled-dimer ground state, where two spins form an interlayer spin singlet. 
 The elementary excitations above this dimerized ground state are gapped triplons, corresponding
 to the breaking of   singlet dimers into spin-1 triplet states. We find that these 
 triplons, which are bosonic quasiparticles with $S=1$, exhibit a nontrivial topological band
 structure, which is characterized by a $\mathbbm{Z}_2$ index, akin to the quantum spin Hall effect~\cite{fu_kane}.
 As a result, the triplons exhibit exotic behaviors, such as a triplon spin Hall effect and counterpropagating triplon edge modes.  
We note that these triplons are different from Refs.~\cite{ss_hall,ss_hall_exp}, where the triplon bands have a Chern index, break time-reversal symmetry and occur in an ordered phase. 
We briefly show that the topological triplons of our example model  occur also in other bilayer systems with strong spin-orbit coupling, such as triangular- or square-lattice bilayer structures~\cite{supp}.
Moreover, this physics also arises in spin-orbit coupled d$^{4}$ Mott insulators~\cite{khaliullin_d4_mott_PRL_13,trivedi_d4_mott_PRB_15}, 
in which spin and orbital moments are canceling each other out.

\begin{figure}
\centering 
\subfloat[]{\includegraphics[width=0.39\textwidth]{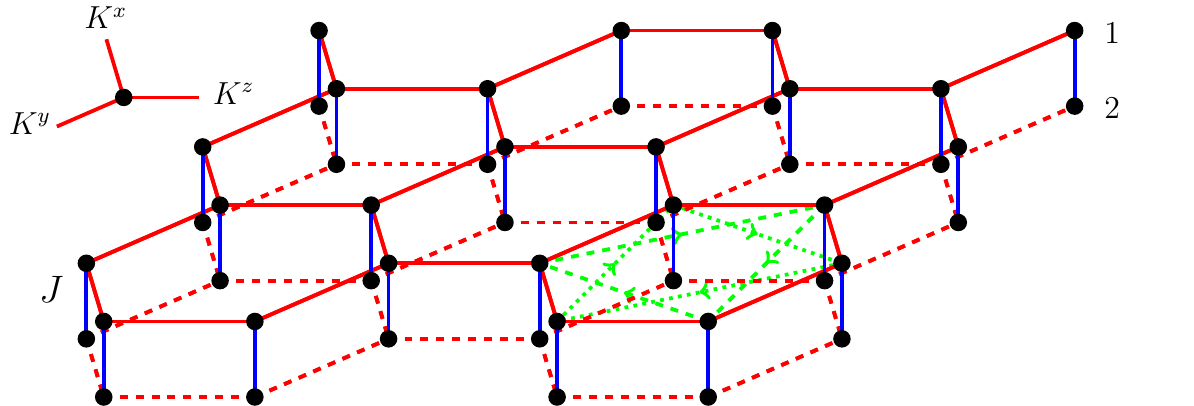}}   
\subfloat[]{\includegraphics[width=0.08\textwidth]{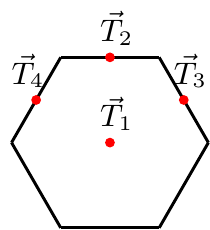}}
\caption{
(a) Honeycomb bilayer model with $S=1/2$ spins indicated by the black dots.
Blue lines represent the interlayer antiferromagnetic interactions, while  
red solid/dashed lines correspond to the anisotropic intralayer exchange.
The DM interaction is perpendicular to the honeycomb layers. Along the green lines, in the direction of an arrow $\dm_{ij}=\dm$ and $\dm_{ji}=-\dm_{ij}$ (shown only on one hexagonal plaquette). 
(b) Hexagonal first Brilliuon zone with the red dots indicating the time-reversal invariant momenta. 
}
\label{fig_model}
\end{figure}

\emph{Model description.--} 
Our model consists of  $S=1/2$ spins on a bilayer honeycomb lattice (Fig.~\ref{fig_model}) with the following Hamiltonian
\begin{align}
\label{eq:ham}
\mathcal{H} &= \sum_{i} \jp_{i} \vec{S}_{1i} \cdot \vec{S}_{2i} 
+ \sum_{\langle ij \rangle} K_{ij} \big[ \vec{S}_{1i} \cdot \vec{S}_{1j} + \vec{S}_{2i} \cdot \vec{S}_{2j} \big] \nonumber \\
&+ \sum_{\langle \langle ij \rangle \rangle} D_{ij} \big[ S^{x}_{1i} S^{y}_{1j} - S^{x}_{1j} S^{y}_{1i} 
+ S^{x}_{2i} S^{y}_{2j} - S^{x}_{2j} S^{y}_{2i} \big] \,,
\end{align}
where $i$ labels the dimer lattice sites and the indices $1,2$ denote the two honeycomb layers.
The first term in Eq.~\eqref{eq:ham} is the antiferromagnetic ($\jp_{i}>0$) interlayer Heisenberg interaction, where we allow for a staggered on-site potential such that $\jp_{i}=\jp \pm \ani$ ($\ani \ll \jp$) on sublattice A (B).  
The second term in Eq.~\eqref{eq:ham} represents the nearest-neighbour Heisenberg interaction within a layer, and the last term is the next-nearest-neighbor  DM interaction. 
Note that we have allowed for anisotropic Heisenberg interactions within a layer [see Fig. \ref{fig_model} (a)] such that, 
$\jn_{ij} = \jn^{\alpha}$ along the $\alpha-$bond ($\alpha=x,y,z$). 
For simplicity, we shall consider $\jnx=\jny\equiv\jn$ such that the interaction $\jnz$ introduces anisotropy, which could be realized in real materials by applying uniaxial pressure. 
We note that the DM interaction is perpendicular to the honeycomb layers such that $\dm_{ij}=\dm (-\dm)$ when going clockwise (anti-clockwise) in a hexagonal plaquette,
see  Fig.~\ref{fig_model}(a).

We are interested in the dimer-paramagnetic phase, described by a product-state of singlets, which is realized for   dominant $\jp >0 $. In this phase
there are three gapped quasiparticle excitation bands, corresponding to the three spin-1 triplet excited states on each dimer
$|t_x\rangle = -[|\uparrow\uparrow\rangle - |\downarrow\downarrow\rangle]/\sqrt{2}$, 
$|t_y\rangle = \ii[|\uparrow\uparrow\rangle + |\downarrow\downarrow\rangle]/\sqrt{2}$, and 
$|t_z\rangle = [|\uparrow\downarrow\rangle + |\downarrow\uparrow\rangle]/\sqrt{2}$, over the singlet state $|t_0\rangle = [|\uparrow\downarrow\rangle - |\downarrow\uparrow\rangle]/\sqrt{2}$. 
To describe the band structure of these triplon excitations we employ 
the bond-operator formalism \cite{ss_bhatt}, 
wherein the triplon quasiparticles are expressed in terms of the triplon creation and annihilation operators
$t_{\gamma}^{\dagger}$ and $t_{\gamma}^{\ }$ ($\gamma=x,y,z$),
defined as $t_{\gamma}^{\dagger} |t_0\rangle = |t_\gamma \rangle$.
Inserting the triplon representation of the spin operators into Eq.~\eqref{eq:ham} yields an interacting triplon Hamiltonian~\cite{supp}. 
For simplicity, we shall work within the harmonic approximation, 
retaining only the bilinear part of the triplon Hamiltonian.
This approximation is justified, since deep inside the paramagnetic phase 
the triplon density is small, which allows to neglect any triplon interactions \cite{ha_note}. 
Within the harmonic approximation, the $t_z$ mode is decoupled from the
$t_x$ and $t_y$ modes.
For that reason, we focus on the $t_x$ and $t_y$ excitations, whose dynamics in 
momentum space in described by
\begin{equation}
\label{eq:ham_k}
\mathcal{H}_{2} = \frac{1}{2} \sum_{\kk} \Psi^{\dagger}_{\kk} \mathcal{M}_{\kk} \Psi_{\kk} \,,
\end{equation}
with the $8 \times 8$ matrix 
\begin{eqnarray} \label{eq_big_M}
\mathcal{M}_{\kk}
&=&
\begin{pmatrix}
h_{1, \kk} & h_{2, \kk} \cr
h^{\dag}_{2, \kk} & h^{T}_{1, -\kk } 
\end{pmatrix}
\end{eqnarray}
and  
$\Psi_{\kk} = \left( t^{A}_{\kk x}, t^{A}_{\kk y}, t^{B}_{\kk x}, t^{B}_{\kk y}, t^{A \dagger}_{-\kk x}, t^{A \dagger}_{-\kk y}, 
t^{B \dagger}_{-\kk x}, t^{B \dagger}_{-\kk y} \right)^{T}$,
where the supercripts $A/B$ label the two honeycomb sublattices.  For now we set the staggered on-site potential $\ani=0$. 
The matrix elements of $\mathcal{M}_{\kk}$ 
\begin{align}
\label{eq:h2h1}
h_{2,\kk} &= h_{1,\kk}-\jp \mathbbm{1} 
= \vec{d} \cdot \vec{\Gamma} \,,
\end{align}
are given in terms of the $\vec{d}$ vector
\begin{align}
\label{eq:dvec}
\vec{d} &= \lbrace \Re(\kappa_{\kk}), -\Im(\kappa_{\kk}), 0, -2\dm \gk, 0 \rbrace \,, 
\end{align}
and the five Dirac matrices
$
\vec{\Gamma} = \lbrace \sigma_1 \otimes \mathbbm{1}, \sigma_2 \otimes \mathbbm{1}, \sigma_3 \otimes \tau_1, \sigma_3 \otimes \tau_2, \sigma_3 \otimes \tau_3 \rbrace
$.
Here $\sigma_{i}$ and $\tau_{i}$ are the Pauli matrices acting on the sub-lattice and the triplon-flavor spaces respectively.
The parameters $\kappa_{\kk}$ and $\gk$ in Eq.~\eqref{eq:dvec} are defined as follows
\begin{align}
\label{eq:kappa} 
\kappa_{\kk} &= \frac{1}{2} \left[ \jnz + \jn e^{\ii \kk_{1}} + \jn e^{\ii \kk_{2}} \right] \,, \\
\label{eq:gk}
\gk &= -\sin(\kk_{1}) + \sin(\kk_{2}) + \sin(\kk_{1}-\kk_{2}) \,,
\end{align}
where $\kk_{1,2} = \kk \cdot \vec{a}_{1,2}$, with $\vec{a}_{1,2} = \lbrace \pm \hat{x}/2, \sqrt{3} \hat{y}/2 \rbrace$ being the Bravais basis vectors. 


\begin{figure*}[th]
\centering
\includegraphics[width=1.0\textwidth]{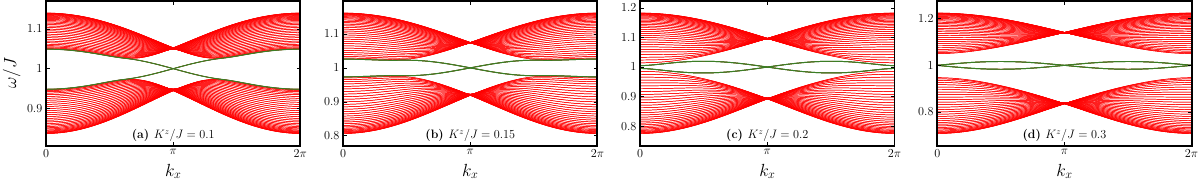}
\caption{ 
Triplon bands in the presence of DM interaction with open boundaries along the zig-zag edge. 
The edge states, which are located around $\jp = 1$, are indicated in green. 
For $|\jnz|<2\jn$, a $\zt$ topological quantum paramagnet is realized, which has similar band structure and edge state as the quantum spin Hall effect. 
At $|\jnz| = 2\jn$ at topological phase transition occurs, which separates the  $\zt$ topological phase
from a quantum paramagmet with edge states that are detached from the bulk bands.
The parameters used in these plots are $\jn/\jp=0.1$, $\dm/\jp=0.01$, and $\ani/\jp=0$.
}
\label{fig_edge_dm}
\end{figure*}

\emph{Triplon dynamics and edge states.--} 
To compute the triplon dispersions, we have to evaluate the eigenvalues of the non-Hermitian matrix $\Sigma \mathcal{M}_{\kk}$, where $\Sigma = \sigma_{3} \otimes \mathbbm{1}_{4 \times 4}$~\cite{diab}. Since 
$\big[ h_{1,\kk}, h_{2,\kk} \big]=0$, the eigenvalues of $\Sigma \mathcal{M}_{\kk}$ are obtained in a straightforward manner. 
For the $t_x$ and $t_y$ triplons the dispersion is
\begin{equation}
\label{eq:wxy}
\omega^{x/y}_{A,B} = \sqrt{\jp \left( \jp \pm 2 \sqrt{4 \dm^{2} \gamma^{'2}_{\kk} + |\kappa_{\kk}|^{2} } \right)} \, ,
\end{equation}
while for the $t_z$-triplon it reads $\omega^{z} = \sqrt{\jp (\jp \pm 2 |\kappa_{\kk}|)}$. 
We now exclusively focus on the $t_x$ and $t_y$ bands, since the $t_z$ band is topologically trivial. 
It is clear from Eq.~\eqref{eq:wxy} that in the absence of the DM interaction and as long as  $|\jnz|<2\jn$, the triplon bands cross each other at two points in the Brillouin zone (BZ).
Since these two band crossings are fourfold degenerate and the dispersion is linear in their vicinity, 
they realize triplon analogues of Dirac fermions, i.e., ``Dirac triplons".
The topological character of these Dirac triplons manifests itself in the edge spectrum
in terms of dispersionless triplon edge states, which connect the two Dirac points,
see Fig.~S1 in the Supplemental Material (SM)~\cite{supp}. 

Upon introducing the DM interaction, a topological gap is opened at the two Dirac points. Two counterpropagating triplon edge states appear within this gap,
which connect the $t_x$ and $t_y$ bulk bands to each other, see Fig.~\ref{fig_edge_dm}. 
These are protected by the time-reversal symmetry.
We may call this state a ``topological triplon insulator", since its edge state spectrum is identical 
to the one of the two-dimensional electronic topological insulator~\cite{hasan_kane}. 
However, as opposed to electronic topological insulators, the edge states of the topological triplon insulator are excited states, which
cross at an energy of the order $J$ above the ground state energy. Hence, in order to probe the physics of these triplon edge states,
they need to be thermally populated or excited out of equilibrium.

\begin{figure}
\centering
\includegraphics[width=0.4\textwidth]{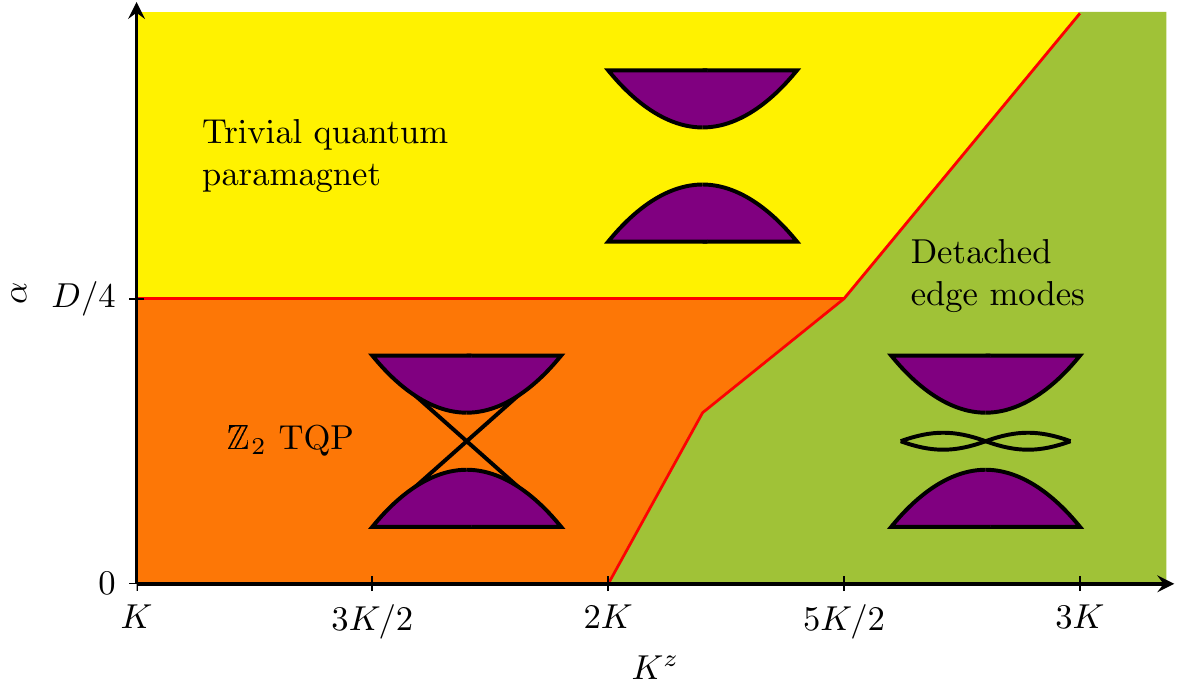}
\caption{Topological phase diagram as a function of the anisotropic Heisenberg interaction $\jnz$ and staggered on-site potential $\ani$ in the presence of a finite DM interaction $\dm$. 
  The lower left phase (orange) is the $\zt$ topological quantum paramagnet or the topological triplon insulator, which has counterpropagating edge modes connecting the bulk triplon modes. The right-side phase (green) is a topological quantum paramagnet with detached edge modes protected by chiral symmetry and realized for stronlgy anisotropic $\jnz$. The upper phase (yellow) is a trivial quantum paramagnet with no edge excitations.  
The parameters used here are $\jn/\jp=0.1$ and $\dm/\jp=0.01$. 
}
\label{fig:pd}
\end{figure}

In Fig.~\ref{fig_edge_dm} we show how the triplon edge states evolve as a function of the anisotropy in the intralayer Heisenberg interaction $K^z/K$.  
Upon increasing $K^z$ relative to $K$, the gapped Dirac triplons move along the edges of the bulk BZ until they merge at the M point for $K^z= 2K$, where they form a quadratic band touching [Fig.~\ref{fig_edge_dm}(c)].
In this process, the crossing of the triplon edge states gets streched out, until at $K^z=2K$ the edge states touch at both $k_x = 0$ and $k_x =\pi$.
Further increasing $K^z / K$, a bulk gap opens up again and the edge states detach from the bulk bands.  
In fact, for $K^z > 2 K$ the triplon edge states lie completely in the bulk  gap, without touching the bulk triplon bands at all [Fig.~\ref{fig_edge_dm}(d)]. 
We will see below that the bulk gap closing at $K^z = 2 K$ corresponds to a topological phase transition, which separates two distinct toplogical phases
with two different types of edge states: For $K^z < 2 K$ there are two edge states attached to the bulk bands and protected by a $\mathbbm{Z}_2$ invariant,
while for $K^z > 2 K$ the edge states are detached from the bulk bands and protected by chiral symmetry. 
In the supplemental material \cite{supp} we also discuss the effect of staggered potential $\ani$. The full phase diagram as a function of both $\ani$ and $\jn^{z}$ is shown in Fig. \ref{fig:pd}. In passing, we remark that these types of detached edge states can be realized, in principle, also in fermionic systems \cite{supp}.

\emph{$\zt$ topological invariant.--} 
In order to establish the topological origin of the edge states discussed above, we show that they are protected by a $\zt$ invariant, which can be defined in the presence of the time-reversal symmetry.
In the presence of parity symmetry (i.e., when $\alpha = 0$),
the $\zt$ invariant $\nu$ can be expressed in terms of the parity eigenvalues of the triplon bands \cite{fu_kane}, i.e., 
\begin{equation}
\label{eq:def_z2}
(-1)^{\nu} = \prod_{i=1}^4 \delta_i \,, 
\end{equation}
where $\delta_i =  \delta ( \vec{T}_i )$ gives the parity eigenvalue of the lower triplon band at the four time-reversal invariant momenta $\vec{T}_{i}$, see Fig.~\ref{fig_model}(c). 
Parity symmetry acts on the triplon Hamiltonian~\eqref{eq:ham_k} as
$\po \mathcal{M}_{\kk} \po^{-1} = \mathcal{M}_{-\kk}$, with the parity operator
$\po = \mathbbm{1} \otimes \Gamma_{1}$.  
Note that for our bosonic problem the relevant matrix 
is not $\mathcal{M}_{\kk}$, but rather
$\Sigma \mathcal{M}_{\kk}$, which, however, obeys the same
parity symmetry as $\mathcal{M}_{\kk}$, since $[ \po, \Sigma ] =0$. 
Recall that for the honeycomb lattice, there are four time-reversal invariant 
momenta: $\vec{T}_{1} = \lbrace 0,0 \rbrace$, $\vec{T}_{2} = \lbrace 0,2\pi / \sqrt{3} \rbrace$, $\vec{T}_{3} = \lbrace \pi, \pi / \sqrt{3} \rbrace$, and $\vec{T}_{4} = \lbrace -\pi, \pi / \sqrt{3} \rbrace$  [see Fig. \ref{fig_model}(b)]. 
At these time-reversal invariant momenta, $ [ \po, \mathcal{M}_{T_{i}}  ]=0$ as well as $ [ \po, \Sigma \mathcal{M}_{T_{i}}  ]=0$.
It is now straightforward to see that the parity eigenvalue of the lower triplon band is related to the sign of $d_{1} = \Re(\kappa_{\kk})$.
Hence, from Eq.~\eqref{eq:kappa} we find that 
\begin{align}
\label{eq:del}
\delta_i 
&= -\mathrm{sgn}\left[ \frac{1}{2} \left( \jnz + 2\jn \cos\left(\frac{k_{x}}{2} \right) \cos\left( \frac{\sqrt{3}k_{y}}{2} \right) \right) \right] \,.
\nonumber
\end{align}
At the four time-reversal invariant momenta we have
\begin{equation}
\label{eq:del_1234}
d_{1}(\vec{T}_{1,2}) = \frac{\jnz \pm 2\jn}{2} \,; ~~~ d_{1}(\vec{T}_{3,4}) =  \frac{\jnz}{2} \,.
\end{equation}
It follows that for $\jn>0$ the $\mathbbm{Z}_2$ invariant is given by 
\begin{equation}
\label{eq:z2}
\nu =
\begin{cases}
0, & \text{if} ~|\jnz|>2\jn \\
1, & \text{if} ~|\jnz|<2\jn \,.
\end{cases}
\end{equation}
Thus, for $|\jnz|<2\jn$ the triplon spectrum is topological, 
in agreement with the appearance of edge states, as discussed above.
Due to its topological excitation spectrum, we call the phase  \mbox{$|\jnz|<2\jn$}
a \emph{$\zt$ topological quantum paramagnet}.

In the absence of parity symmetry (i.e., when $\alpha \ne 0$), formula~\eqref{eq:def_z2} for the $\mathbbm{Z}_2$ topological invariant $\nu$
is no longer valid. However, in this case we can  consider an adiabatic deformation, which smoothly transforms
the Hamiltonian to a parity symmetric one, without closing the gap in the triplon spectrum. 
Alternatively, the $\mathbbm{Z}_2$ invariant can be formulated in terms of the triplon eigenstates~\cite{km2,fu_kane_PRB_06},
which does not rely on the existence of a parity symmetry, but only the time-reversal symmetry.

Another point to note is that the $\zt$ invariant is independent of the DM interaction. The role of the DM interaction is to just separate the triplon bulk bands. In fact, just 
as in the case of the quantum spin-Hall effect, absence of DM interaction simply means that the triplon bands must touch at energy $\jp$ for $|\jnz|<2\jn$ \cite{fu_kane}. 
 
 
 
\emph{Protection of detached edge modes.--}
The $\mathbbm{Z}_2$ invariant~\eqref{eq:def_z2} is zero for $|\jnz|>2\jn$.
Nevertheless, in this regime there appear edge states too, which are protected by symmetry, as we will now show.
To establish this, we first observe that besides the antiunitary time-reversal symmetry, the triplon Hamiltonian $\mathcal{M}_{\kk}$, Eq.~\eqref{eq_big_M}, also exhibits
a unitary symmetry, i.e., it commutes with  $\hat{\mathcal{G}} = \mathbbm{1}_{4 \times 4} \otimes \tau_2$.
Moreover, the model possesses a type of chiral symmetry, that is $ \mathcal{M}_{\kk} - \jp \mathbbm{1}$ anticommutes with 
the chiral operator $\ch = \mathbbm{1}   \otimes \sigma_3 \otimes \tau_3$. Since both of these
symmetries also hold for $\Sigma \mathcal{M}_{\kk}$, 
it follows that: (i) the triplon bands can be labelled by the eigenvalues of $\hat{\mathcal{G}}$ and
(ii) the triplon bands are symmetric around the energy~$J$. 

We can investigate the edge-state wavefunction near $k_x=0$ by making an ansatz: $\Psi_{ed}(y) = e^{-\lambda y} \Phi$. It turns out that $\lambda =(\jnz - 2\jn)/\sqrt{3}\jn$, which means that these edge modes exists only when $\jnz>2\jn$ (green region in Fig.~\ref{fig:pd}). We refer to the supplemental material \cite{supp} for technical details. We quote here the full wavefunction  of the detached edge states,
\begin{equation}
\label{eq:ed_wf}
\Phi_{1,2} = \left( u \phi_{1,2} , v \phi_{1,2} \right)^{T} \,,
\end{equation}
where $u^{2} - v^{2} =1$, which follows from the bosonic bogoliubov transformation and  $\phi_{1,2}=\left( \varphi_{\pm}, 0 \right)^{T}$, with $\varphi_\pm$ the eigenfunctions of  $\tau_{2}$ (i.e., $\tau_2 \varphi_{\pm} = \pm \varphi_{\pm}$).

From Eq.~\eqref{eq:ed_wf} we infer that the two edge states $\Phi_{1,2}$ have
opposite eigenvalues with respect to $\hat{\mathcal{G}}$, i.e., $\hat{\mathcal{G}} \Phi_{1,2} = \pm \Phi_{1,2}$.
Hence, since $\hat{\mathcal{G}}$ commutes with $\Sigma \mathcal{M}_{\vec{k}}$, any hybridization between the two edge modes 
is prohibited by the symmetry $\hat{\mathcal{G}}$.
Moreover, we find that chiral symmetry $\hat{\mathcal{C}}$ 
converts one edge state into another (i.e., $\hat{\mathcal{C}} \Phi_{1,2} = - \Phi_{2,1}$)
and that parity $\hat{P}$ guarantees the degeneracy between states
on opposite edges.
Therefore, away from $k_x = 0$
there is always exactly one edge state with $\epsilon  < \jp$  
and one edge state with $\epsilon  > \jp$, and hence 
the $k_x =0$ band crossing is pinned at $\epsilon = J$~\footnote{In the absence of parity (i.e., for $\alpha \ne 0$), the edge states on opposite edges split in energy, but are still symmetric
around $J$. }.   
Furthermore, we observe that as we let $k_x \to - k_x$ the $\hat{\mathcal{G}}$ eigenvalues
of the $\epsilon_{1,2}$ eigenstates get interchanged. For this reason
and due to the $2 \pi$ periodicity of the wavefuncitons, there must be another
crossing  of the edge states between $0$ and $\pi$. 
Due to time-reversal symmetry this second crossing is pinned at $k_x = \pi$. 

%



\emph{Conclusions and implications for experiments.--}
In this paper, we have presented a spin model on a honeycomb bilayer,
which exhibits topological triplon excitations protected by  time-reversal symmetry. 
We have shown that two of the three triplon excitation bands  carry a nontrivial
$\mathbbm{Z}_2$ number.
By the bulk-boundary correspondence, this  leads to two 
counterpropagating triplon edge modes with helical dispersions, similar to the quantum spin Hall insulator. 
Furthermore, we have shown that upon making one of the Heisenberg couplings stronger,
the spin system undergoes a topological phase transition into
a phase, where the counterpropagating edge modes are completely detached from the bulk excitations. 

These triplon edge modes could potentially be used as robust and efficient channels for spin transport.
Their topological origin protects them against disorder scattering. Triplon-triplon scattering 
is weak, because of the dilute density of triplons, provided we are away from a magnetic quantum critical point. 
Moreover,  the interaction-induced damping of the triplon edge modes is suppressed,
 due to the absence of Goldstone modes and due to phase-space constraints, as long as the bulk triplon gap is larger than $\jp/2$. 
The triplon edge modes should be observable in various experimental probes. 
For example, neutron-scattering experiments should be able to detect a pronounced peak in the dynamical
spin structure factor at the energy of the triplon edge states (see SM~\cite{supp} for a detailed prediction).
Another possibility is to measure spin-Hall noise in a normal metal deposited on top of the honeycomb bilayer paramagnet,
which is expected to show signatures of the triplon edge states~\cite{shns}. Apart from this, thermal or spin transport measurements could also probe these non-trivial edge modes \cite{ruckriegel_PRB_18}. However, unlike fermionic topological states there is no quantized response, 
which makes it challenging to find an unambiguous physical observable of the nontrivial topology.

The topological triplon edge states discussed in this paper are expected to occur 
in a wide range of model systems and materials.  
A promising set of materials is that of Chromium trihalides, CrX$_3$ (X= F, Cl, Br, I), which are layered honeycomb materials with relevant interlayer coupling. These were of great interest in the past as a prototypical example of Heisenberg ferromagnets \cite{prb3,prb4}, but have since then been forgotten. However, recently they have been shown to host Dirac magnons \cite{prx2018}. It might be possible to realize a singlet ground state in these systems with the application of external pressure or by substituting Chromium with some other transition metal. 
Moreover, the physics discussed in this work is
  expected to exist also in other bilayer systems with strong spin-orbit coupling, such as triangular- or square-lattice bilayer structures
with dimerized ground states~\cite{supp}. This  might be  of relevance  (after appropriate substitution) for BaCuSi$_2$O$_6$~\cite{PhysRevB.55.8357}, which exhibits
a spin-singlet dimerized ground state in a square-lattice bilayer structure.

Topological triplons can also arise in various spin-orbital systems realizing singlet-triplet phenomenon. 
In particular, $d^4$ Mott insulators such as  Li$_2$RuO$_3$~\cite{miura_Li2RuO3_JPSJ_07} and Ag$_3$LiRu$_2$O$_6$~\cite{argyriou_Ag3LiRu2O6}, where transition metal ions form a honeycomb lattice, are promising candidates as they display singlet-triplet physics  due to strong spin-orbit coupling \cite{khaliullin_d4_mott_PRL_13,trivedi_d4_mott_PRB_15}. It will be interesting to work out the conditions under which
the triplon edge states can arise  in these $d^4$ Mott insulators. 
It may be extended to $d^8$ Mott insulators as well \cite{gchen}.
Another direction  for future research is the study of
magnetic quantum phase transitions  from a topological quantum paramagnet to a magnetically ordered phase. 
There are indications that the ordered phase might also host topological edge excitations \cite{katsura}.
However, this might require exact numerical studies.


\emph{Acknowledgment.--
}We thank G.\ Khaliullin and H.\ Takagi for useful discussions.



\bibliographystyle{apsrev}
\bibliography{bib_z2_tqp_v1}

\begin{thebibliography}{46}
\bibitem[{ha_()}]{ha_note}
\bibinfo{note}{In general, the harmonic approximation is controlled in large
  dimensions, both in the paramagnetic as well as inside the magnetically
  ordered phases, as shown in Refs.~\cite{larged_para,larged_af}.}
\end{thebibliography}


 \clearpage

\appendix

\setcounter{equation}{0}
\setcounter{figure}{0}
\setcounter{table}{0}
\setcounter{page}{1}
\renewcommand{\theequation}{S\arabic{equation}}
\renewcommand{\thefigure}{S\arabic{figure}}
\renewcommand{\bibnumfmt}[1]{[S#1]}
\renewcommand{\citenumfont}[1]{S#1}

\onecolumngrid
\begin{center}
\textbf{\large Supplemental Material: \\
\smallskip
$\mathbbm{Z}_{2}$ topological quantum paramagnet on a honeycomb bilayer} \\
\medskip 

Darshan G. Joshi and Andreas P. Schnyder \\

\smallskip

{\em Max-Planck-Institute for Solid State Research, D-70569 Stuttgart, Germany}

\end{center}

\smallskip

\twocolumngrid

In this supplemental material we give 
the derivation of the triplon Hamiltonian, present the triplon spectrum
in the absence of DM interactions and in the presence of the staggered potential $\alpha$, and compute the dynamical structure factor.
We also briefly discuss the triplon excitations in a square-lattice bilayer model.

\section{I.~Bond-operator Hamiltonian}

In this section we discuss the triplon Hamiltonian used in the main text. At each dimer singlet there are three spin-1 excitations which are described using bosonic quasiparticles, triplons. It is then possible to express the spin operators in terms of these triplon annihilation and creation operators using the bond-operator theory~\cite{ss_bhatt_s} as follows: 
\begin{align}
\label{eq:s_op}
S^{\alpha}_{1,2i} &= \frac{1}{2} \big[ t^{\dagger}_{i\alpha} P_{i} \pm P_{i} t_{i\alpha} 
- \ii \epsilon_{\alpha \beta \gamma} t^{\dagger}_{i\beta} t_{i\gamma} \big] \,, \\
\label{eq:s1s2}
\vec{S}_{1i} \cdot \vec{S}_{2i} &= -\frac{3}{4} + \sum_{\alpha} t^{\dagger}_{i\alpha} t_{i\alpha} \,.
\end{align}
In the above equations, $t_{\alpha}$ ($\alpha=x,y,z$) is the triplon annihilation operator. Since the triplons are bosonic in nature one needs to introduce a hard-core constraint, i.e., no more than one boson per dimer. This is addressed via the projection operator, $P_{i}=1-\sum_{\alpha} t^{\dagger}_{i\alpha} t_{i\alpha}$ \cite{collins_s}, which eliminates matrix elements between the physical and unphysical states in the Hilbert space. It is now straightforward to obtain the triplon Hamiltonian by inserting Eqs. \eqref{eq:s_op} and \eqref{eq:s1s2} into the spin Hamiltonian [Eq.~(1)] in the main text:
\begin{align}
\mathcal{H} &= -\frac{3}{4} \jp N + \sum_{i \alpha} \jp_{i} t^{\dagger}_{i\alpha} t_{i\alpha}  \nonumber \\
&+ \sum_{\langle ij \rangle,\alpha} \frac{\jn_{ij}}{2} \big[ t^{\dagger}_{i\alpha} P_{i} P_{j} t_{j \alpha}
+ t^{\dagger}_{i\alpha} P_{i} t^{\dagger}_{j\alpha} P_{j} + H.c. \big] \nonumber \\
&+ \sum_{\langle \langle ij \rangle \rangle} \frac{\dm_{ij}}{2} \big[ t^{\dagger}_{ix} P_{i} P_{j} t_{j y} + t^{\dagger}_{ix} P_{i} t^{\dagger}_{jy} P_{j}  
\nonumber \\
&~~~~~~~~~~- t^{\dagger}_{iy} P_{i} P_{j} t_{j x} - t^{\dagger}_{iy} P_{i} t^{\dagger}_{jx} P_{j} + H.c. \big] \,,
\label{eq:ham_r}
\end{align}
where, $N$ is number of dimer sites. As is evident, we have now obtained an interacting triplon Hamiltonian. However, the situation can be simplified by resorting to the {\em harmonic approximation}, i.e., considering only those pieces in the Hamiltonian which are bilinear in triplon operators. Such an approximation is well justified when the density of triplons is very small such that the triplon interaction terms are insignificant. The density of triplons is very small ($\lesssim 20 \%$) away from a magnetic quantum critical point, the situation we are indeed dealing with. Moreover, it has been shown that the harmonic approximation is well controlled in large dimensions, such that corrections beyond it can be arranged in a systematic expansion in inverse spatial dimension \cite{larged_para_s, larged_af_s}. Hence we can safely ignore the triplon interaction terms as the physics discussed here will not be changed qualitatively. The bilinear triplon Hamiltonian is then given as follows:
\begin{align}
\mathcal{H}_{2} &= \sum_{i \alpha} \jp_{i} t^{\dagger}_{i\alpha} t_{i\alpha}  
+ \sum_{\langle ij \rangle, \alpha} \frac{\jn_{ij}}{2} \big[ t^{\dagger}_{i\alpha} t_{j \alpha}
+ t^{\dagger}_{i\alpha} t^{\dagger}_{j\alpha}  + H.c. \big] \nonumber \\
&+ \sum_{\langle \langle ij \rangle \rangle} \frac{\dm_{ij}}{2} \big[ t^{\dagger}_{ix} t_{j y} + t^{\dagger}_{ix} t^{\dagger}_{jy} 
- t^{\dagger}_{iy} t_{jx} - t^{\dagger}_{iy} t^{\dagger}_{jx}  + H.c. \big] \,.
\label{eq:ham_har}
\end{align}
It is easy to see that at the harmonic level, the $t_{z}$ mode does not interact with the other two triplon modes. Using the lattice translation symmetry we can Fourier transform the above bilinear Hamiltonian to obtain the quadratic Hamiltonian in momentum space (Eq.~(3) in the main text).

\begin{figure*}[t]
\centering
\includegraphics[width=1.0\textwidth]{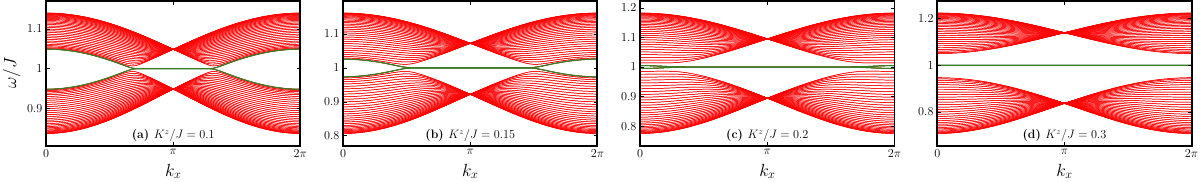}
\caption{ 
Triplon bands in the absence of the DM interaction with open boundaries along the zig-zag edge. 
The edge states are located at $\jp = 1$. Upon increasing $|\jnz|$, the band-touching points move and {\em gap} 
out via quadratic band touching at $|\jnz|=2\jn$. The edge states, however, survive even for larger $|\jnz|$.
The parameters used here are $\jn/\jp=0.1$ and $\dm/\jp=\ani/\jp=0$.
}
\label{fig_edge_dm0}
\end{figure*}

In order to see the topological edge modes we need to solve for the energymodes of Eq. \eqref{eq:ham_har} (a non-Hermitian eigenvalue problem \cite{diab_s}) with periodic boundary condition along the $x$ direction, i.e., the direction running parallel to the zig-zag edge, while keeping the zig-zag boundaries open. Thus we obtain energy spectra as a function of $k_x$ as plotted in Fig.~(2) in the main text as well as Figs.~\ref{fig_edge_dm0} and \ref{fig_edge_lam}. In the topological phase we can find edge modes located along the zig-zag edges.

\section{II.~Absence of spin-orbit coupling}

It is clear that in the absence of the DM interaction the triplon spectrum has {\em gapless} points, i.e. the two bands touch each other, where $|\kappa_{\kk}|=0$. It is easy to see that 
this is always true as long as $|\jnz|<2\jn$ and that the dispersion around these points is linear. 
Note that when $|\jnz|=\jn$, the {\em gapless} points are located at the corners of the Brilliuon zone. Upon increasing $|\jnz|$ compared to $\jn$, these band-touching points move 
along the edges of the Brillouin zone and at $|\jnz|=2\jn$ these points merge at the $M$ point forming a quadratic band touching point in $k_x$ direction. Eventually, when 
$|\jnz|>2\jn$ the two bands again do not touch each other. Consequently, there is a flat zig-zag edge mode at energy $\jp$. This is shown in Fig. \ref{fig_edge_dm0}. On the other hand, upon decreasing $|\jnz|$ compared to $K$, the band-touching points move inside the Brilliuon zone and 
eventually at $|\jnz|=0$ there is a line of band touching along $k_x=\pi$.

Analogously in graphene without spin-orbit coupling, there will be a flat zero-energy mode completely detached from the bulk once one of the hopping parameters is greater than twice the other hoppings. This might have potential application in dissipationless transport. 

 
\section{III.~Effect of staggered potential}

Note that in the absence of the DM interaction, a staggered potential ($\ani$) simply opens a bulk gap and the edge states are dispersionless connecting only the respective bands. In other words, the edge modes do not cross each other within the band gap and hence it is a trivial scenario wherein these edge modes could be adiabatically pushed into the bulk modes. This is 
similar to the case of adding a trivial mass to the graphene.

\begin{figure*}[t]
\centering
\includegraphics[width=1.0\textwidth]{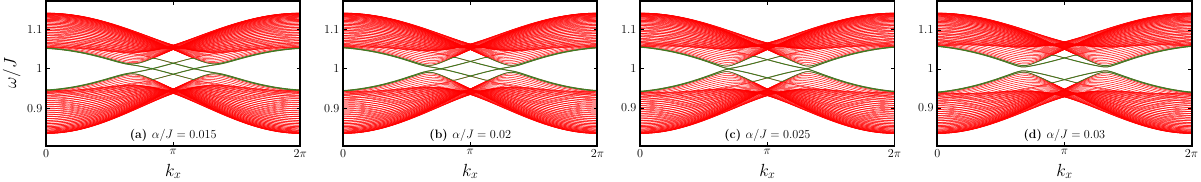}
\caption{ \label{fig_edge_lam}
Triplon bands in the presence of DM interaction and staggered potential $\ani$, with open boundaries along the zig-zag edge. 
The edge states are indicated in green. 
For small $\ani$, a $\zt$ topological quantum paramagnet is realized, which has similar band structure and edge states as the quantum spin Hall insulator. 
Since a finite $\ani$ breaks parity symmetry, the triplon states on opposite zig-zag edges have different dispersions.
Upon increasing $\ani$, the  bulk gap closes  at $\ani = D/4$.
For $\alpha >   D/4$ 
we go into a trivial quantum paramagnetic phase with gapped edge states. The parameters used here are $\jn/\jp=\jn^{z}/\jp=0.1$ and $\dm/\jp=0.01$.
}
\end{figure*}

As discussed in the main text, a DM interaction opens a topological gap between the bulk modes in the absence of a staggered potential. The resulting edge modes are protected by a $\zt$ invariant. Upon adding a small staggered potential these topological edge modes still survive as long as the triplon band gap does not close. However, since a non-zero $\ani$ breaks parity symmetry, the edge modes are split (see Fig.~\ref{fig_edge_lam}). Upon increasing $\ani$ the band gap first reduces such that eventually at $\ani=\dm/4$ the band gap closes and the system becomes trivial upon reopening of the gap for $\ani > \dm/4$. This is shown in Fig.~\ref{fig_edge_lam}. 

As stated earlier, even for a small fixed $\ani$ in presence of a DM interaction a $\zt$ topological quantum paramagnet is realized. In such a case, upon increasing $\jnz$ for a fixed $\ani$ and $\dm$ we once again have a topological phase transition to obtain detached edge modes. However, in presence of $\ani$ there is no simple analytic expression to obtain this phase transition point. Nevertheless we can obtain it numerically. A full phase diagram as a function of $\ani$ and $\jnz$ in presence of a finite DM interaction is shown in Fig.~(3) in the main text.

\section{IV.~Protection of detached edge modes}

To determine the symmetry properties of the detached edge states,
we now derive the explicit form of the edge-state wavefunction in the vicinity of $k_x =0$.
For that purpose, we set $\alpha =0$ and expand the triplon Hamiltonian $\mathcal{M}_{\kk}$, Eq.~(3) in the main text, up to linear order around $\kk=\lbrace 0, 2\pi/\sqrt{3} \rbrace$,
where the gap between the two triplon bands is minimal. 
Since the normal and the anomalous parts of $\mathcal{M}_{\kk}$ share the same eigenspace (because $\big[ h_{1,\kk}, h_{2,\kk} \big]=0$),
the edge modes of the Hamiltonian can be constructed from the edge state solutions of the normal piece, $h_{1} (\kk )$.
The mode at the zig-zag edge $y=0$ can thus be determined
by solving the equation $h_1 ( k_x, -\ii \partial_{y} ) \psi_{ed} = \epsilon \psi_{ed}$,
with the ansatz $\psi_{ed}(y) = e^{-\lambda y} \phi$. Note that the ansatz $\psi_{ed}$
decays exponentially into the bulk, with inverse decay length $\lambda > 0$.
With this, the secular equation $\textrm{det} [  h_1 (k_x, + \ii \lambda ) - \epsilon \mathbbm{1}_{4 \times 4}] = 0$ 
yields
\begin{equation}
\label{eq:edge1}
\begin{vmatrix}
d'_{4} \tau_2 + (\jp-\epsilon) \mathbbm{1} & (d'_{1} + \lambda d'_{2}) \mathbbm{1} \\
(d'_{1} - \lambda d'_{2}) \mathbbm{1} & -d'_{4} \tau_2 + (\jp-\epsilon) \mathbbm{1} 
\end{vmatrix}
 = 0 \, ,
\end{equation}
where  $d'_{1}=(\jnz-2\jn)/2$, $d'_{2}=\sqrt{3}\jn/2$, and $d'_{4}=-4\dm k_{x}$. 
Solving Eq.~\eqref{eq:edge1} for $\lambda$, we obtain
\begin{equation}
\label{eq:lam}
%
\lambda_\pm =  \pm \frac{1}{d'_{2}}  \sqrt{ d'^{2}_{4} + d'^{2}_{1} - (\jp-\epsilon)^{2} } \,.
\end{equation}
For the decaying solution $\lambda_+$,
the eigenstates are $\phi_{1,2}=\left( \varphi_{\pm}, 0 \right)^{T}$, with $\varphi_\pm$ the eigenfunctions of  $\tau_{2}$ (i.e., $\tau_2 \varphi_{\pm} = \pm \varphi_{\pm}$),
and the edge-state energies are  $\epsilon_{1,2} = \jp \pm d'_{4}$. 
It then follows that $\lambda_+ =(\jnz - 2\jn)/\sqrt{3}\jn$, from which it is  clear that   an edge 
state exists only when $\jnz>2\jn$ (green region in Fig.~(3) in the main text). 
Using the eigenstates $\phi_{1,2}$, we can now construct
the full wavefunction  of the detached edge states
\begin{equation}
\label{eq:ed_wf}
\Phi_{1,2} = \left( u \phi_{1,2} , v \phi_{1,2} \right)^{T} \,,
\end{equation}
where $u^{2} - v^{2} =1$, which follows from the bosonic bogoliubov transformation.
The corresponding energies are  $\epsilon_{1,2} = \sqrt{\jp^{2} \pm 2 \jp d'_{4} }$.

\section{V.~Dynamical structure factor}

\begin{figure}
\centering 
\subfloat[]{\includegraphics[width=0.25\textwidth]{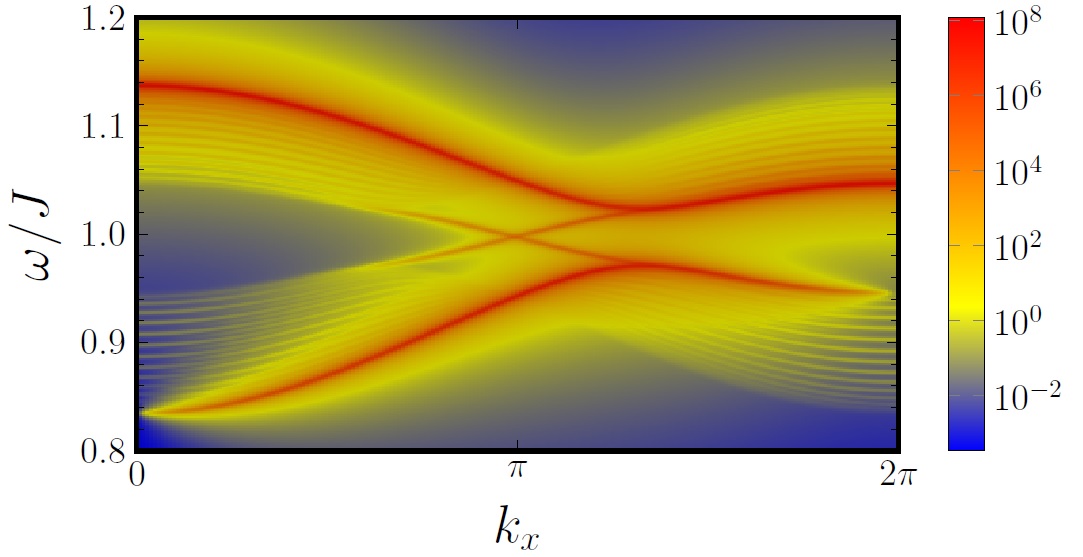}} ~~
\subfloat[]{\includegraphics[width=0.25\textwidth]{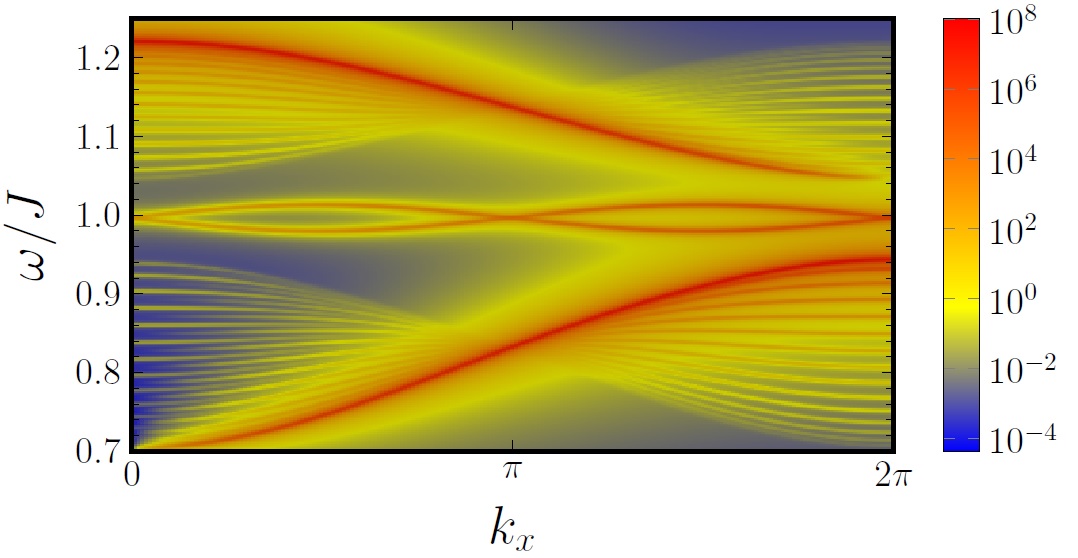}}
\caption{Dynamical structure factor obtained by summing Eqs.~\eqref{eq:sx_def} and \eqref{eq:sy_def}. Note that we have included a Lorentzian broadening $\delta/\jp=10^{-3}$ in these plots. Plot (a) and (b) correspond to Figs.~2(a) and 2(d) in the main text, respectively. }
\label{fig:sfct_k}
\end{figure}

For spin systems dynamical structure factor is an important observable, which is accessible in neutron scattering experiments.  
The dynamical structure factor is given as follows:
\begin{equation}
\label{eq:sf_def}
S(\kk,\omega) = \frac{1}{N} \sum_{i,j} S_{ij} e^{\ii \kk \cdot \vec{r}_{ij}} \,.
\end{equation}
Here $S_{ij} = - \Im \chi_{ij}$, with $\chi$ being the spin correlation function and $\Im$ stands for the imaginary part.
In order to detect the edge modes we calculate this quantity with zig-zag edges in the $y-$direction while retaining lattice-translation symmetry in the $x-$direction. Thus $k_x$ is still a good quantum number. In an experiment, such a quantity will correspond to $k_y=0$ component.
So,
\begin{equation}
\label{eq:s_w}
S(k_x,\omega) \equiv S(k_y=0,\omega) = \frac{1}{N} \sum_{i,j} S_{ij} e^{\ii k_{x} \hat{k}_{x} \cdot \vec{r}_{ij}} \,.
\end{equation}
There are two channels for spin-spin correlations, even and odd, owing to the two layers. This is calculated with respect to
$\vec{S}_{1} \pm \vec{S}_{2}$ for the even and odd channel respectively. Using the spin expressions in terms of triplons, it is then straightforward to see that within the single-mode approximation (no continuum contribution) 
only the odd channel is relevant. To proceed further, we need to express the triplon opertors in terms of operators corresponding to the Bogoliubov quasiparticle, which diagonlize the triplon Hamiltonian. This is as follows:
\begin{align}
\label{eq:tax}
t^{A}_{x,i} &= \sum^{4N}_{m=1} \big[ u_{i,m} \tau_{m} + v_{i,m} \tau^{\dagger}_{m} \big] \,, \\
\label{eq:tay}
t^{A}_{y,i} &= \sum^{4N}_{m=1} \big[ u_{i+N,m} \tau_{m} + v_{i+N,m} \tau^{\dagger}_{m} \big] \,, \\
\label{eq:tbx}
t^{B}_{x,i} &= \sum^{4N}_{m=1} \big[ u_{i+2N,m} \tau_{m} + v_{i+2N,m} \tau^{\dagger}_{m} \big] \,, \\
\label{eq:tby}
t^{B}_{y,i} &= \sum^{4N}_{m=1} \big[ u_{i+3N,m} \tau_{m} + v_{i+3N,m} \tau^{\dagger}_{m} \big] \,.
\end{align}
Here, $u_{i,m}$ and $v_{i,m}$ are Bogoliubov coefficients which are elements of $4N \times 4N$ matrices U and V respectively. Combination of these matrices diagonalizes the triplon Hamiltonian. The $\tau$ operators correspond to the Bogoliubov quasiparticles. 
We can thus calculate the odd channel contribution to the dynamical structure factor, with the following contributions:
\begin{widetext}
\begin{align}
\label{eq:sx_def}
S^{xx}_{ij} &= \sum^{4N}_{m=1}  \delta(\omega - \omega_{m}) 
\bigg[ v^{*}_{i,m} u^{*}_{j,m} + v^{*}_{i,m} v_{j,m} 
+ u_{i,m} u^{*}_{j,m} + u_{i,m} v_{j,m}  \nonumber \\
&~~~~~~~~+ v^{*}_{i,m} u^{*}_{j+2N,m} + v^{*}_{i,m} v_{j+2N,m} 
+ u_{i,m} u^{*}_{j+2N,m} + u_{i,m} v_{j+2N,m} \nonumber \\
&~~~~~~~~+ v^{*}_{i+2N,m} u^{*}_{j,m} + v^{*}_{i+2N,m} v_{j,m} 
+ u_{i+2N,m} u^{*}_{j,m} + u_{i+2N,m} v_{j,m} \nonumber \\
&~~~~~~~~+ v^{*}_{i+2N,m} u^{*}_{j+2N,m} + v^{*}_{i+2N,m} v_{j+2N,m} 
+ u_{i+2N,m} u^{*}_{j+2N,m} + u_{i+2N,m} v_{j+2N,m} 
\bigg] \,, \\
\label{eq:sy_def}
S^{yy}_{ij} &= \sum^{4N}_{m=1} \delta(\omega - \omega_{m}) 
\bigg[ v^{*}_{i+N,m} u^{*}_{j+N,m} + v^{*}_{i+N,m} v_{j+N,m} 
+ u_{i+N,m} u^{*}_{j+N,m} + u_{i+N,m} v_{j+N,m}  \nonumber \\
&~~~~~~~~+ v^{*}_{i+N,m} u^{*}_{j+3N,m} + v^{*}_{i+N,m} v_{j+3N,m} 
+ u_{i+N,m} u^{*}_{j+3N,m} + u_{i+N,m} v_{j+3N,m} \nonumber \\
&~~~~~~~~+ v^{*}_{i+3N,m} u^{*}_{j+N,m} + v^{*}_{i+3N,m} v_{j+N,m} 
+ u_{i+3N,m} u^{*}_{j+N,m} + u_{i+3N,m} v_{j+N,m} \nonumber \\
&~~~~~~~~+ v^{*}_{i+3N,m} u^{*}_{j+3N,m} + v^{*}_{i+3N,m} v_{j+3N,m} 
+ u_{i+3N,m} u^{*}_{j+3N,m} + u_{i+3N,m} v_{j+3N,m} 
\bigg]  \,.
\end{align}
\end{widetext}
Here $\omega_{m}$ are $4N$ eigenmodes of the triplon Hamiltonian constructed for $N$ zig-zag stripes stacked along $y$ direction. 
Summing the above two contributions we obtain the dynamical structure factor. This is shown is Fig.~\ref{fig:sfct_k}, where we have also included a Lorentzian broadening $\delta/\jp=10^{-3}$. The edge modes are clearly seen around the energy $\omega=\jp$ in both the cases, $\zt$ topological quantum paramagnet [Fig.~\ref{fig:sfct_k} (a)] as well as the case with detached edge modes [Fig.~\ref{fig:sfct_k} (b)].

\begin{figure}[t!]
\centering 
\includegraphics[width=0.49\textwidth]{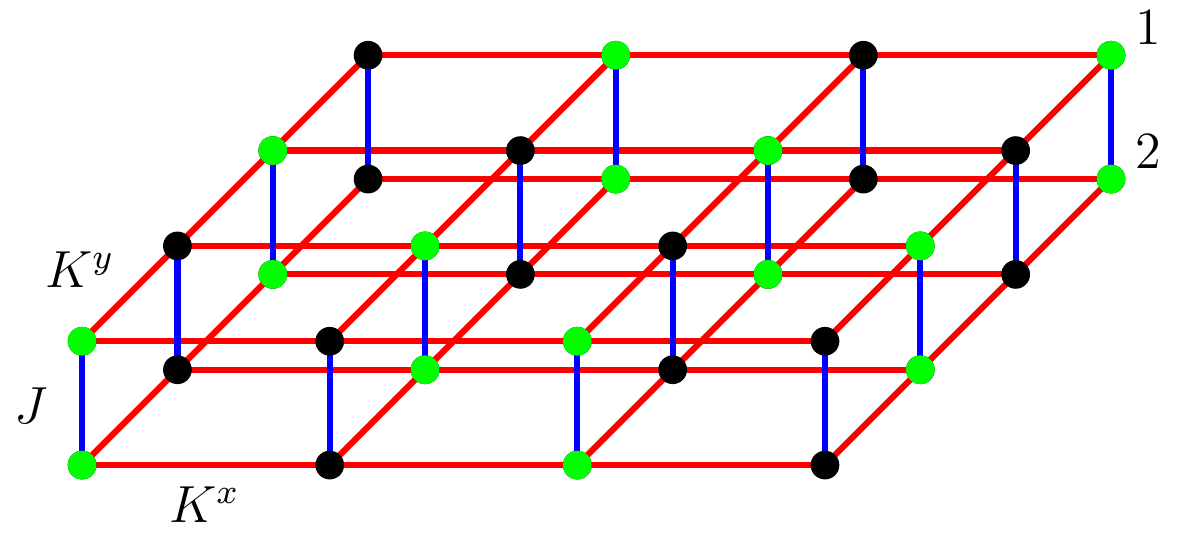}
\caption{Square-lattice bilayer with two dimer sublattices (black and green sites) which could host a {\em topological triplon insulator}.  }
\label{fig:square_new}
\end{figure}

\section{IV.~Square-lattice bilayer model}

The $\zt$ topological quantum paramagnet  (also known as the {\em topological triplon insulator}) discussed in the main text can be realized 
not only on bilayer honeycomb lattices, but also 
on other bilayer systems.  The main ingredient to realize this exotic state is to have two independent pairs of degrees of freedom. In the main text these were the two triplon flavors and the two sublattices of the honeycomb lattice. In general, any bipartite bilayer lattice  with time-reversal symmetry, but broken spin-rotation symmetry, has these ingredients.
Moreover, a bilayer lattice with two orbitals per lattice site (as realized, e.g., in $d^4$ Mott insulators) could also provide the necessary ingredients.

For example, a square lattice bilayer could  host the model described in the main text. However, since there is only one atom in the unit cell, we are in shortage of a degree of freedom. This can be overcome if every site also has two orbital degrees of freedom, or alternatively, if one considers a bilayer square lattice with two atoms per unit cell, as shown in Fig. \ref{fig:square_new}. Such a bilayer-lattice system can be viewed as two interpenetrating bilayer square lattices. We hope that this could be realized in real materials by substituting alternate sites of the original bilayer square lattice with different ions/atoms. Perhaps controlled substitution can be made on the well studied square-lattice bilayer compound BaCuSi$_2$O$_6$.

\section{VI.~Fermionic systems}

 A fermionic tight-binding Hamiltonian is easily obtained by replacing the bosonic operators with fermionic operators in the bosonic hopping Hamiltonian (Eq. (5) in the main text). In fact, the resultant tight-binding Hamiltonian describes the physics of graphene with spin-orbit coupling, as considered by Kane and Mele \cite{km1_s,km2_s}. The counter-part of the anisotropic Heisenberg interaction ($\jn^{z}$) considered here is an anisotropic hopping in the case of graphene. As a consequence, graphene-like systems with  strongly anisotropic hopping realize completely detached zig-zag edge-states around zero energy, which could have application in dissipationless transport. 
These detached edge states could be created, for example,  in synthetic anisotropic honeycomb lattices of nanofabricated semiconductor structures~\cite{wang_artificial_graphene_Nat_Nano_18_s}. 
Similar physics has been recently discussed in the context of phosphorene~\cite{phosp1_s,phosp2_s}.

\end{document}